\documentclass[aps,preprint,nofootinbib]{revtex4}
\begin{document}

\title{On the 150th anniversary of Maxwell equations} 

\author{Alexei M. Frolov}
\email{afrolov@uwo.ca}

\affiliation{Department of Applied Mathematics, University of Western Ontario, N6A 5B7, London,
             Canada }


\begin{abstract}
In this lecture we discuss some interesting developments in the modern theory of electromagnetic field(s). In particular, 
by using the methods developed in Dirac's constraint dynamics we derive the Schr\"{o}dinger equation for the free 
electromagnetic field. The electromagnetic field that arises only contains combinations of transverse photons and does 
not include any scalar and/or longitudinal photons. This approach is also used to determine and investigate the actual 
symmetry of the free electromagnetic field. Then we discuss the Majorana representation of the Maxwell equations, the 
symmetric form of Maxwell equations and an original approach to electrodynamics called the ``scalar electrodynamics''. \\

\noindent
DOI: 10.13140/2.1.4634.4484 \\
PACS number(s): 04.20.Fy \\

\noindent
(A substantial part of this manuscript was originally presented in my public lecture, entitled: ``150 years with Maxwell 
equations'', presented on 27th of September 2012 at Queen's University, Kingston, Ontario, Canada) \\

\noindent 
First version 20.10.2014, Preprint-2014-15/1 (this is 2nd version) [electromagnetic waves;optics], 20 pages.

\end{abstract}

\date{\today}

\maketitle

\newpage

\section{Introduction}

This paper is a detailed version of my public lecture which was dedicated to an important anivesary for the whole modern physics. Indeed, one hundred and
fifty years ago Maxwell presented his famous equations \cite{Maxw} in a closed and relatively simple form. These equations described results of various 
electric and magnetic experiments known at that time (see, e.g., \cite{BW} and references therein). Moreover, it followed directly from those equations 
that all electric and magnetic phenomena can be explained with the use of the united `electro-magnetic field'. In the general case, the corresponding 
vectors $({\bf E}, {\bf H})$ include six components each of which is a function of the spatial (Cartesian) coordinates ${\bf r}$ and time $t$. The 
propagation of the free electromagnetic fields in four-dimensional space-time continuum was discovered later by Hertz (theoretically) and has become a most 
crucial part of today's human communications. Step-by-step, Maxwell equations started to be applied to very large spectrum of phenomena many of which were 
not originally considered as being electric, or magnetic. These equations successfully survived two great events in physics of the last century: (a) the 
Galilean-Lorentz crisis around 1900 - 1905, and (b) the appearance of Quantum Mechanics around 1925 - 1927. Presently the whole Maxwell theory of 
electro-magnetic phenomena is considered to be a solid and absolute construction in modern physics. The age of this theory is also outstanding and it is in 
many dozens times larger than an average `time-life' of 99.99 \% of the `fundamental' theories being developed today.

In this brief lecture I want to show that our understanding of the Maxwell equations is still a very interesting and evolving area of theoretical 
physics. In particular, by following Dirac we will see how Maxwell equations survived another crucial event in the physics of XX century - a transition to 
another mechanics which was created by Dirac in 1951 for Hamiltonian systems with constraints and now it is known as the `constraint dynamics'. 
We also discuss the dynamical symmetry of the free electromagnetic filed(s), the Majorana form of the Maxwell equations and the co-existence of our `electric' 
and alternative `magnetic' worlds. These two worlds can be located at the same spatial place, but cannot interect with each other directly. The only
possible communication between our `real' and alternative `ghost' worlds comes in the lowest-order approximation in the fine structure constant 
$\alpha = \frac{e^2}{\hbar c} \approx \frac{1}{137}$ due to radiation. Finally, we discuss so-called scalar electrodynamics - a new approach to 
electrodynamics which is based on the using only four scalar functions.

In general, a detailed description of the time-evolution of various physical systems and fields is a fundamental problem which arises in many areas of physics. 
Explicit derivation of the equations which govern the time-evolution of physical systems and fields is the most interesting part of physics. In Quantum 
Electrodynamics (QED) the time-evolution of the electromagnetic field (or, EM-field, for short) is governed by the Schr\"{o}dinger (or Heisenberg) equations 
for each of the field components. For the free EM-field(s) such an equation is known since the end of 1920's \cite{One} - \cite{Three}. Later, analogous 
equations were obtained for arbitrary electromagnetic fields which interact with electrons and positrons \cite{Heitl}, \cite{AB}. By solving such equations 
people answered a significant number of questions which arose in Quantum Electrodynamics. However, the main disadvantage of these (Schr\"{o}dinger) equations 
was a presence of indefinite numbers of the scalar and longitudinal photons. In general, the presence of arbitrary numbers of scalar and longitudinal photons 
transforms all QED-calculations into extremely painful process, which in many cases does not lead to a consistent answer. To avoid complications related to the 
presence of large (even infinite) numbers of the scalar and longitudinal photons scientists working in this area developed quite a number of clever tricks 
and procedures. The most recent and widely accepted approach is based on exact compensation of the scalar photons by an equal number of longitudinal photons. 
All such procedures, however, are not based on an internal logic of the original QED-theory of the EM-field(s). 

In 1950's Dirac developed his famous mechanics \cite{Dirac0}, \cite{Dirac1} of the constrained dynamical systems with Hamiltonians. By applying this mechanics 
to the free electromagnetic field one finds that it produces the EM-field which is represented as a linear combination of only transverse photons and does not 
include any scalar and/or longitudinal photons. In other words, such a EM-field can directly be used in QED calculations to determine the probabilities of 
different processes. Briefly, we can say that Dirac constraint dynamics (or Dirac mechanics) allows one to produce electromagnetic field which is physical, i.e. 
it does not include any of the ghost components. Therefore, we can investigate the properties of such a field and assume that this field coincides with the 
actual (or physical) EM-field. In particular, we can determine the actual symmetry of the equations of the free electromagnetic field.  

\section{Hamiltonian}

Following Dirac \cite{Dirac1} we begin our analysis from the Lagrangian $L$ of the free electromagnetic field written 
in the Heaviside-Lorentz units (see, e.g., \cite{LLE})
\begin{equation}
 L = -\frac14 \int F_{\mu\nu} F^{\mu\nu} dx dy dz = -\frac14 \int F_{\mu\nu} F^{\mu\nu} d^3x \; \; \; , \; \; 
 \label{eq1}
\end{equation}
where the integration is over three-dimensional space and $F_{\mu\nu}$ and $F^{\mu\nu}$ are the covariant and 
contravariant components of the $F-$tensor which is uniformly related with the corresponding derivatives of the field 
potential $A_{\mu}$ (or $A^{\nu}$) by the relations
\begin{equation}
 F_{\mu\nu} = A_{\nu,\mu} - A_{\mu,\nu} \; \; \; , \; \; \label{eq2}
\end{equation}
where the suffix with the comma before it means differentiation according to the following general scheme
$T_{,\mu} = \frac{d T}{d x^{\mu}}$, where $T$ is an arbitrary quantity (or tensor) and $x = (x^{0}, x^{1}, x^{2}, 
x^{3})$ is the point in the four-dimensional space-time. Note that the suffix `0' with the comma before it 
designates the temporal derivative (or time derivative), while analogous notations with suffixes 1, 2 and 3 mean 
the corresponding spatial derivatives. 

Following Dirac \cite{Dirac1} we need to construct the Hamiltonian of the free electromagnetic field by using the 
Lagrangian $L$ from Eq.(\ref{eq1}). First, by varying the corresponding velocities, i.e. temporal components of the 
tensor $F$, we introduce the momenta $B^{\mu}$   
\begin{equation}
 \delta L = - \frac12 \int F^{\mu\nu} \delta F_{\mu\nu} d^3x = \int F^{\mu0} \delta A_{\mu,0} d^3x = 
 \int B^{\mu} \delta A_{\mu,0} d^3x \; \; \; , \; \; \label{eq3}
\end{equation}
As follows from Eq.(\ref{eq3}) the momenta $B^{\mu}$ are defined by the equalities $B^{\mu} = F^{\mu0} = -F^{0\mu}$,
which follow from Eq.(\ref{eq3}), and antisymmetry of the $F-$tensor, i.e. from $F^{\mu\nu} = -F^{\nu\mu}$ (see, 
e.g.,  \cite{Heitl}, \cite{Dirac1} and \cite{LLE}). From this definition of the momenta one finds that $B^{0}$ equals 
zero identically, since $B^{0} = F^{00} = -F^{00} = - B^{0}$. This is the primary constraint which is designated in 
Dirac's constrained dynamics as $B^{0} \approx 0$. In Quantum Elelctrodynamics this can be written in the more 
informative form $B^{0} \Psi = 0$ (or $B^{0} \mid \Psi \rangle = 0$), where $\Psi$ (also $\mid \Psi \rangle$) is the wave 
function of the free electromagnetic field. Briefly, this means that for all states of the free electromagnetic field 
which are of interest for our purposes below we have $B^{0} \Psi = 0$, or $B^{0} \mid \Psi \rangle = 0$. 

Now, we can construct the Hamiltonian of the free electromagnetic field, or EM-field, for short. It should be mentioned here
that any Hamiltonian determines a simplectic structure with the dimension $2 n + 1$, where $2 n$ is the number of dynamical 
variables, i.e. $n$ coordinates and $n$ momenta conjugate to these coordinates. For the free electromagnetic field in 
three-dimensional space we have four generalized coordinates $A_{\mu} = (A_0, A_1, A_2, A_3)$ of the field, or four-vector 
$(\phi, {\bf A})$ of the field potentials in the traditional $EM-$notations. The momenta $B^{\mu}$ conjugate to these 
coordinates also form 4-vector $(B^{0}, B^{1}, B^{2}, B^{3})$. The Poisson brackets between these dynamical variables must 
be equal to the delta-function, i.e.
\begin{equation}
 [ B^{\mu}(x), A_{\nu}(x^{\prime})] = - [ A_{\nu}(x), B^{\mu}(x^{\prime})] = -g^{\mu}_{\nu} \delta^{3}(x - x^{\prime}) \label{eq4}
\end{equation}
All other Poisson brackets between these dynamical variables, i.e. the $[B^{\mu}(x), B^{\nu}(x^{\prime})]$ and $[A_{\mu}(x), 
A_{\nu}(x^{\prime})]$ brackets, equal zero identically. 
  
By using the Lagrangian, Eq.(\ref{eq1}), and explicit formulas for the momenta $B^{\mu} = F^{\mu0} = -F^{0\mu}$ we can obtain 
the explicit expression for the Hamiltonian $H$. The first step here is to write the Hamiltonian in terms of the velocities
($A_{\mu,0}$ and $F_{r0}$): 
\begin{equation}
 H = \int B^{\mu} A_{\mu,0} d^3x - L =  
 \int ( F^{r0} A_{r,0} + \frac14 F^{rs} F_{rs} + \frac12 F^{r0} F_{r0} ) d^3x \; \; \; , \; \; \label{eq5}
\end{equation}
where the indexes $r$ and $s$ stand for the spatial indexes, i.e. $r$ = 1, 2, 3, and $s$ = 1, 2, 3. For the first term in 
the second equation we can write $A_{r,0} = F_{0r} - A_{0,r} = - F_{r0} - A_{0,r}$ (this follows from the definition of 
$F_{\mu\nu}$, Eq.(\ref{eq2})). This allows one to transform the Hamiltonian, Eq.(\ref{eq5}), to the form 
\begin{equation}
 H = \int (\frac14 F^{rs} F_{rs} - \frac12 F^{r0} F_{r0} + F^{r0} A_{0,r}) d^3x = 
 \int \Bigl(\frac14 F^{rs} F_{rs} + \frac12 B^{r} B^{r} - (B^{r})_{,r} A_{0}\Bigr) d^3x \; \; \; , \; \; \label{eq51}
\end{equation}  
where we introduce the momenta $B^{r}$ and integrated the last term ($F^{r0} A_{0,r}$) by parts. This is the explicit formula for 
the Hamiltonian $H$ of the free electromagnetic field. Let us investigate this Hamiltonain $H$, Eq.(\ref{eq51}). First, it is easy 
to see that the Poisson bracket of the momentum $B^{0}$ and the Hamiltonian $H$ (i.e. $[B^{0}, H]$) equals $(B^{r})_{,r} \delta^3(x - 
x^{\prime})$. In Dirac's constrained dynamics the Poisson brackets between the primary constraints and Hamiltonian determine the 
secondary constraints. In other words, the secondary constraint for the free electromagnetic field equals $(B^{r})_{,r}$, i.e. to 
the sum of spatial derivatives of the corresponding components of the momenta $B^{r}$. In three-dimensional notations this value 
equals to $div {\bf B}$. 

By determining the Poisson bracket between the secondary constraint $(B^{r})_{r}$ and the Hamiltonian, Eq.(\ref{eq51}), one finds 
that it equals zero identically. This means that Dirac's procedure is closed, since no (non-zero) tertiary constraints have been 
found. The final expression for the total Hamiltonian $H_T$ of the electromagnetic field is
\begin{equation}
 H_T = H + \int v(x_1, x_2, x_3) B^{0} d^3x = 
 \int \Bigl(\frac14 F^{rs} F_{rs} + \frac12 B^{r} B^{r} - A_{0} (B^{r})_{r} + v B^{0} \Bigr) d^3x \; \; \; , \; \; \label{eq52}
\end{equation}
where $v = v(x_1, x_2, x_3)$ is an arbitrary coefficient defined in each point of three-dimensional space. This Hamiltonian is a `classical' 
expression. Our next goal is to quantize and obtain the quantum Hamiltonian operator which corresponds to the classical Hamiltonian, 
Eq.(\ref{eq52}). This problem is considered in the next Section.    
 
\section{Quantization} 

The total Hamiltonian $H_T$, Eq.(\ref{eq52}), derived above allows one to perform the quantization of the free electromagnetic field and derive
the Schr\"{o}dinger equation which describes time-evolution of the EM-field. The general process of quantization for various classical systems 
with Hamiltonians is described in detail in various textbooks (see, e.g, \cite{GT}, \cite{Dir} and \cite{Fock}). Briefly, such a process of 
quantization can be represented as a following two-step procedure. The first step is the replacement of the classical fields by the corresponding
quantum operators. The classical Poisson bracket, Eq.(\ref{eq4}), is replaced by the quantum Poisson bracket in which all classical momenta are 
replaced by the differential operators. The quantum Poisson bracket for two operators of the electromagnetic field must include the reduced Plank 
constant $\hbar = \frac{h}{2 \pi}$ and, may be, the speed of light in vacuum $c$. The presence of the speed of light in the expressions for Poisson 
brackets depends upon the explicit form of the field operators and the units used. For the operators $B^{\mu}(x)$ and $A_{\nu}(x^{\prime})$ 
defined above the transformation from the classical to the quantum Poisson bracket is written in the form
\begin{equation}
 [ B^{\mu}(x), A_{\nu}(x^{\prime})]_{C} = -g^{\mu}_{\nu} \delta^{3}(x - x^{\prime}) \rightarrow 
 [ B^{\mu}(x), A_{\nu}(x^{\prime})]_{Q} = \hbar (-g^{\mu}_{\nu}) \delta^{3}(x - x^{\prime}) \; \; \; , \; \; \label{eq45}
\end{equation}
where $B^{\mu}(x)$ and $A_{\nu}(x^{\prime})$ are the operators and $B^{\mu}(x)$ is the differential operator in the $A_{\nu}(x)$-representation (or 
coordinate representation). Other notations in Eq.(\ref{eq45}) have the same meaning as in Eq.(\ref{eq4}). The second step of the quantization 
process is the explicit introduction of the wave function $\Psi$ which depends upon time $t$ and all coordinates of the dynamical system, i.e. 
upon the $A_{\mu} = (A_0, A_1, A_2, A_3)$ components of the electromagnetic field, i.e. $\Psi = \Psi(A_0, A_1, A_2, A_3)$. The Hamiltonian and other 
`observable' quantities must now be considered as operators which act (or operate) on such wave functions. At this point we have to introduce the
system of traditional notations for different components of the electromagnetic field and their derivatives. The four-vector potential of the 
electromagnetic field is represented as the unique combination of its scalar component $A_0$, which is usually designated as $\phi$, and three 
remaining components, which form a three-dimensional vector ${\bf A} = (A_1, A_2, A_3) = (A_x, A_y, A_z)$ (see, e.g., \cite{Heitl} and references 
therein). The wave function of the electromagnetic field $\Psi$ is a function of the scalar $\phi$ and vector $\bf{A}$, i.e. $\Psi = \Psi(\phi, {\bf 
A})$. 

As follows from the definition of momenta of the free electromagnetic field ($B^{\mu} = F^{\mu0} = -F^{0\mu}$) such momenta essentially coincide with the 
corresponding components of the electric field ${\bf E}$, i.e. $B^{\mu} = -F^{0\mu} = E^{\mu} = - E_{\mu}$. On the other hand, as follows from 
Eq.(\ref{eq45}) the same momenta can be considered as differential operators in the $A_{\nu}(x)$-representation, or coordinate representation. In  other 
words, we can also choose the following definition of the momenta $B^{\mu}(x) = -\frac{\partial}{\partial A_{\nu}(x)}$, or $B^{\mu}(x) = -\hbar 
\frac{\partial}{\partial A_{\nu}(x)}$ in the case of quantum Poisson brackets. For general Hamiltonian systems such a twofold representation of momenta 
are acceptable, since transition from one to another does not change the fundamental Poisson brackets and, therefore, does not lead to any noticeable 
contradiction with the reality and/or with the first principles of the Hamiltonian approach. Now, we can write for the primary constraint
\begin{equation}
 - \hbar \frac{\partial}{\partial \phi} \mid \Psi(\phi, {\bf A}) \rangle = 0
\end{equation}
This means that the wave function $\mid \Psi \rangle$ of the free electromagnetic field cannot depend upon the scalar component (or $\phi-$component), 
i.e. $\mid \Psi \rangle = \Psi(A_1, A_2, A_3) = \Psi(A_x, A_y, A_z)$, where $A_x, A_y$ and $A_z$ are the three Cartesian coordinates of the vector 
${\bf A}$. 

An arbitrary three-dimensional vector ${\bf A} = (A_x, A_y, A_z)$ can always be represented (see, e.g., \cite{Kochin}) as a linear combination of its 
longitudinal $A_{\parallel}$ and two transverse $A^{(1)}_{\perp}, A^{(2)}_{\perp}$ components, i.e. ${\bf A} = (A_x, A_y, A_z) = (A_{\parallel}, 
A^{(1)}_{\perp}, A^{(2)}_{\perp})$. By using the standard methods of vector analysis (see, e.g., \cite{Kochin}) it can be shown that the condition
$div {\bf A} = 0$ in each spatial point is equivalent to the equality $A_{\parallel} = 0$ which must be obeyed in each spatial point. Now, the secondary 
constraint is written in the form
\begin{equation}
 - \hbar \frac{\partial}{\partial A_{\parallel}} \mid \Psi({\bf A}) \rangle = 0
\end{equation}
which leads to the conclusion that the vector $\mid \Psi({\bf A}) \rangle$ depends upon the two transverse components ($A^{(1)}_{\perp}, A^{(2)}_{\perp}$)
only. In other words, for the free electromagnetic field only those states (or wave functions) are acceptable for which $\mid \Psi \rangle = \mid 
\Psi(A^{(1)}_{\perp}, A^{(2)}_{\perp}) \rangle$. Formally, for the free electromagnetic field one can use only such spatial vectors which have only two 
components (at arbitrary time $t$). Moreover, since ${\bf E} = -\frac{1}{c} \frac{\partial {\bf A}}{\partial t}$, then the vector of electric field ${\bf 
E}$ also has the two spatial components only. To simplify our notation below, we shall assume that electromagnetic wave always propagates into $z-$direction 
(in each spatial point) and it has two non-zero components ($x-$ and $y-$components). This means that $\mid \Psi \rangle = \mid \Psi(A_{x}, A_{y}) \rangle$ 
and $A_z = A_{\parallel} = 0$. This important result will be used below.  

The knowledge of the Hamiltonian $H$ written in the canonical variables of `momenta' ${\bf E}$ and `coordinates' ${\bf A}$ of the electromagnetic field 
allows one to obtain all equation(s) of the time-evolution of the free electromagnetic field. In reality, there is an additional problem here related with 
the fact that the Hamiltonian contains only special combinations of spatial derivatives of the coordinates, i.e. $curl {\bf A}$, rather than coordinates 
${\bf A} = (A_x, A_y, A_z)$ themselves. This problem is solved by considering the spatial Fourier transform of the `coordinates', or components of the vector 
${\bf A}$. To simplify analysis even further we represent the original Fourier transform in a `discrete' form, i.e. as an infinite sum, 
e.g., 
\begin{equation}
 {\bf A} =  \sum_{{\bf k} \alpha} \Bigl( c_{{\bf k} \alpha} {\bf A}_{{\bf k} \alpha} +  c^{*}_{{\bf k} \alpha} {\bf A}^{*}_{{\bf k} \alpha} \Bigr) = 
 \sum_{{\bf k} \alpha} \sqrt{\frac{c^2}{2 \omega}} \Bigl[ {\bf e}^{(\alpha)} \exp(\imath {\bf k} \cdot {\bf r}) c_{{\bf k} \alpha} + 
 {\bf e}^{(\alpha)*} \exp(-\imath {\bf k} \cdot {\bf r}) c^{*}_{{\bf k} \alpha} \Bigr]
\end{equation}
where ${\bf A}_{{\bf k} \alpha} = {\bf e}^{(\alpha)} \exp(\imath {\bf k} \cdot {\bf r}) \sqrt{\frac{c^2}{2 \omega}}$ are the normalized plane waves (in the 
Heaviside-Lorentz units), $\omega = c \mid {\bf k} \mid$ and $ {\bf e}^{(\alpha)} \cdot {\bf e}^{(\beta)*} = \delta_{\alpha\beta}$. Analogous plane-wave 
expansions for the electric ${\bf E}$ and magnetic ${\bf H}$ fields are
\begin{equation}
 {\bf E} =  \sum_{{\bf k} \alpha} \Bigl( c_{{\bf k} \alpha} {\bf E}_{{\bf k} \alpha} +  c^{*}_{{\bf k} \alpha} {\bf E}^{*}_{{\bf k} \alpha} \Bigr) \; \; \; , 
  \; \; \;
 {\bf H} =  \sum_{{\bf k} \alpha} \Bigl( d_{{\bf k} \alpha} {\bf H}_{{\bf k} \alpha} +  d^{*}_{{\bf k} \alpha} {\bf H}^{*}_{{\bf k} \alpha} \Bigr)
\end{equation}
where ${\bf E}_{{\bf k} \alpha} = \imath \omega {\bf A}_{{\bf k} \alpha}$ and ${\bf H}_{{\bf k} \alpha} = \imath \omega ({\bf n} \times {\bf A}_{{\bf k} \alpha})$,
The amplitudes $c_{{\bf k} \alpha}, d_{{\bf k} \alpha}$ and their complex conjugate $c^{*}_{{\bf k} \alpha}, d^{*}_{{\bf k} \alpha}$ in such expansions are now 
considered as a canonical (Hamiltonian) variables. Sometimes it is more convenient to introduce the new canonical variables which are the linear combinations of 
the $c_{{\bf k} \alpha}$ and $c^{*}_{{\bf k} \alpha}$ amplitudes and $d_{{\bf k} \alpha}$ and $d^{*}_{{\bf k} \alpha}$. The only non-trivial Poisson bracket is 
$[ c_{{\bf k} \alpha}, c^{*}_{{\bf k} \alpha}] = 1$ and $[ d_{{\bf k} \alpha}, d^{*}_{{\bf k} \alpha}] = 1$   (for classical amplitudes), or 
$[ c_{{\bf k} \alpha}, c^{\dagger}_{{\bf k} \alpha}] = \hbar$ and $[ d_{{\bf k} \alpha}, d^{\dagger}_{{\bf k} \alpha}] = \hbar$ (in the case of quantum amplitudes 
when $c^{*}_{{\bf k} \alpha} \rightarrow c^{\dagger}_{{\bf k} \alpha}$ and analogously for $d-$amplitudes). All other Poisson brackets equal zero 
identically. Note that the both Hamiltonian and Poisson brackets are the quadratic expressions in the Fourier amplitudes of the free electromagnetic field. Therefore, 
it is possible to re-define these amplitudes in the quantum case (by multiplying them by a factor $\frac{1}{\sqrt{\hbar}}$). After such a re-definition the Poisson 
brackets between quantum and classical amplitudes look identically, but the normalized plane waves take an additional factor $\sqrt{\hbar}$, i.e. we must write now: 
${\bf A}_{{\bf k} \alpha} = {\bf e}^{(\alpha)} \exp(\imath {\bf k} \cdot {\bf r}) \sqrt{\frac{\hbar c^2}{2 \omega}}$. Such a representation has a number of advantages 
in applications, since in this case the operators $c^{\dagger}_{{\bf k} \alpha}$ and $c_{{\bf k} \alpha}$ are dimensionless, i.e. they act on the number of photons 
(with the two possible polarizations) in the field (or photon) wave function. In respect with this, the whole procedure of quantization of the amplitudes of the 
Fourier expansion is called the second quantization. 

Finally, the Hamiltonian of the free electromagnetic field (in this case $c_{{\bf k} \alpha} = d_{{\bf k} \alpha}$) is reduced to the infinite sum of Hamiltonians of 
independent harmonic oscillators
\begin{equation}
 H = \sum_{{\bf k} \alpha} \frac12 \hbar \omega \Bigl( c^{\dagger}_{{\bf k} \alpha} c_{{\bf k} \alpha} + c_{{\bf k} \alpha} c^{\dagger}_{{\bf k} \alpha} \Bigr)
 \label{cc*}
\end{equation}
where for each spatial vector ${\bf k}$ one finds two independent harmonic oscillators (for $\alpha = +1$ and $\beta = -1$, or $\alpha = 1$ and $\beta = 2$). Note
that the operators $c^{\dagger}_{{\bf k} \alpha}$ and $c_{{\bf k} \alpha}$ in the last equation are dimensionless, i.e. they act on the total number of photons 
only. All such transformations are described in \cite{LLQ} and here we do not want to repeat them. Note only that the Hamiltonian approach for the free 
electromagnetic field leads to the well known Planck formula for the thermal energy distribution of electromagnetic radiation. This indicates the correctness of
our Hamiltonian $H$, Eq.(\ref{cc*}), derived above. This Hamiltonian is used to solve a large number of actual problems, e.g., we apply this Hamiltonian below to
determine the dynamical symmtery of the free electromagnetic field.  

\section{On the dynamical symmetry of the free electromagnetic field}

The Hamiltonian of the free electromagnetic field, Eq.(\ref{cc*}), is reduced to the form 
\begin{equation}
 H = \sum_{{\bf k} \alpha} \hbar \omega \Bigl( c^{\dagger}_{{\bf k} \alpha} c_{{\bf k} \alpha} + \frac12 \Bigr) = \sum_{{\bf k}} \hbar \omega ( a^{\dagger}_{1}({\bf k}) 
 a_{1}({\bf k}) + a^{\dagger}_{2}({\bf k}) a_{2}({\bf k}) + 1 ) \label{cc1}
\end{equation}
where $a_{1}({\bf k}) =  c_{{\bf k} \alpha} = a_{1}$ and $a_{2}({\bf k}) =  c_{{\bf k} \beta} = a_{2}$. For any given ${\bf k}$ we can write 
\begin{equation}
 H_{{\bf k}} =  \hbar \omega ( a^{\dagger}_{1} a_{1} + a^{\dagger}_{2} a_{2} + 1 ) \label{cc2}
\end{equation}
An interesting question is to determine the symmetry of the corresponding Schr\"{o}dinger equation with the Hamiltonian, Eq.(\ref{cc2}). To answer this question let us
construct the four following operators: $A^{i}_{j} = a^{\dagger}_{i} a_{j}$ which commute with the Hamiltonian, Eq.(\ref{cc2}). The operator $A = \sum_{i} a^{\dagger}_{i} 
a_{i} = \frac{1}{\hbar \omega} H_{{\bf k}} - 1$ also commute with $H_{{\bf k}}$. The commutation relations between $A^{i}_{j}$ operators are:
\begin{equation}
  [ A^{i}_{j}, A^{k}_{l} ] = \delta^{k}_{j} A^{i}_{l} - \delta^{i}_{l} A^{k}_{j}
\end{equation}
These commutation relations coincide with the well known relations between four generators of the $U(2)-$group (the group of unitary $2 \times 2$ matrixes). Now, we 
introduce three $B^{i}_{j}$ operators defined by the relations $B^{i}_{j} = A^{i}_{j} - \frac12 \delta^{i}_{j} A$. Note that for these operators the condition $B^{1}_{1} + 
B^{2}_{2} = 0$ is always obeyed. The three operators $B^{i}_{j}$ are the generators of the $SU(2)$-group, i.e. the group of unitary $2 \times 2$ matrixes and determinants 
of these matrixes equal unity). Thus, the group of dynamical symmetry of the free electromagnetic field is the three-parameter $SU(2)-$group. The physical representations 
of this group which are only of interest in applications are $D(p,q) = D(n,0)$, where $p \ge q$ are non-negative integer and $n = p + q$. Note that the total number of 
parameters in this $SU(2)-$group coincides with the total number of Stokes parameters. 

\section{Maxwell equations and wave propagation}

In all Sections above we ignored an obvious (or internal) relation between Maxwell equations and wave propagation. The nature of this relation directly follows from the 
Hamiltonian approach mentioned in Section II. Indeed, we can write the following expression for the Hamiltonian $H$ which acts on the wave function of the free electromagnetic 
field
\begin{equation}
 H \mid \Psi \rangle = \frac12 \int \Bigl[{\bf E}^2 + (curl {\bf A})^2 \Bigr] d^3x \mid \Psi \rangle = \frac12 \int \Bigl[ \frac{1}{c^2} \Bigl(\frac{\partial {\bf A}}{\partial 
 t}\Bigl)^2 + (curl {\bf A})^2 \Bigr] d^3x \mid \Psi \rangle
\end{equation}
where it is assumed that $\phi = 0$ and $div {\bf A} = 0$ (or $div {\bf E} = 0$) in each spatial point. These two conditions must be written in the form of the following 
equations for the wave function ($\Psi$) $\phi \mid \Psi \rangle = 0$ and $div {\bf A} \mid \Psi \rangle = 0$. From here one finds
\begin{eqnarray}
 \langle \Psi \mid (\delta H) \mid \Psi \rangle &=& \int (\delta {\bf A}) \Bigl[- \frac{1}{c^2} \frac{\partial^2 {\bf A}}{\partial t^2} - 
 curl (curl {\bf A}) \Bigr] d^3x \mid \Psi \rangle  \\
 &=& \langle \Psi \mid \int (\delta {\bf A}) \Bigl[- \frac{1}{c^2} \frac{\partial^2 {\bf A}}{\partial t^2} + \Delta {\bf A} \Bigr] d^3x \mid \Psi 
 \rangle = 0 \nonumber
\end{eqnarray}
This is the energy functional of the free electromagnetic field $H({\bf A})$. 

The condition that the variational derivative $\frac{\delta H}{\delta {\bf A}}$ equals zero in each spatial point leads to the following equation (wave equation) for each 
component of the vector-potential ${\bf A}$: 
\begin{equation}
 (\frac{1}{c^2} \frac{\partial^2 {\bf A}}{\partial t^2} - \Delta {\bf A}) \mid \Psi \rangle = 0 \; \; \;, \; \; \; or \; \; \;  
 \Bigl( \frac{1}{c^2} \frac{\partial^2 {\bf A}}{\partial t^2} - \frac{\partial^2 {\bf A}}{\partial x^2} -
 \frac{\partial^2 {\bf A}}{\partial y^2} - \frac{\partial^2 {\bf A}}{\partial z^2} \Bigr) \mid \Psi \rangle = 0 \label{waveeq}
\end{equation}
This is the Schr\"{o}dinger equation of the free electromagnetic field. Briefly, we can say that the minimum of the functional $H({\bf A})$ (also called the energy functional 
of the free electromagnetic field) uniformly leads to the wave equations for the vector ${\bf A}$ and, therefore, for the vectors of the electric ${\bf E}$ and magnetic ${\bf 
H}$ fields, respectively. The equation, Eq.(\ref{waveeq}), can be written in the form of one of the Maxwell equations:
\begin{equation}
 curl{\bf H} = \frac{1}{c} \frac{\partial {\bf E}}{\partial t}
\end{equation}
Another Maxwell equation follows from the definition of momentum of the free electromagnetic field ${\bf E} = - \frac{1}{c} \frac{\partial {\bf A}}{\partial t}$. By calculating 
$curl$ of both sides of this equation one finds
\begin{equation}
 curl{\bf E} = - \frac{1}{c} \frac{\partial {\bf H}}{\partial t}
\end{equation}
where all these equations must be considered on the wave functions $\mid \Psi \rangle$ for which the condition $(div {\bf E} ) \mid \Psi \rangle = 0$ is obeyed in each spatial point. 
This secondary constraint coincides with another Maxwell equation. The last (fourth) Maxwell equation directly follows from the definition ${\bf H} = curl{\bf A}$. It should be 
mentioned that in modern literature on constraint dynamics the role of constraints is often considered as relatively minor. For electrodynamics of the free electromagnetic field it is 
not true and the constraint $(div {\bf E} ) \mid \Psi \rangle = 0$ allows one to determine many important features of the propagating electromagnetic field. This question is discussed 
in the Appendix. Below, we briefly consider some special form of Maxwell equations known as Majorana representation.

\section{Majorana representation}

Let us discuss another form of Maxwell equations which has a number of advantages in some applications. In this form (obtained first by Majorana) the system of Maxwell equations is 
represented in the form of Dirac equation(s) for masseless particle. To simplify our discussion below we consider the case of the free electromagnetic field. Let us introduce the 
two new vectors ${\bf F} = {\bf E} + \imath {\bf H}$ and ${\bf G} = {\bf E} - \imath {\bf H}$, where $\imath$ is the imaginary unit, and the gradient vector ${\bf p} = 
\Bigl( \frac{\partial}{\partial x}, \frac{\partial}{\partial y}, \frac{\partial}{\partial z} \Bigr) = \nabla$. In these notations Maxwell equations of the free electromagnetic 
field(s) are written in the following form
\begin{eqnarray}
 \frac{1}{c} \frac{\partial {\bf F}}{\partial t} = ({\bf s} \cdot {\bf p}) {\bf F} \; \; \; , \; \; \; 
 {\bf p} \cdot {\bf F} = 0 \label{Max1} \\
 \frac{1}{c} \frac{\partial {\bf G}}{\partial t} = ({\bf s} \cdot {\bf p}) {\bf G} \; \; \; , \; \; \; 
 {\bf p} \cdot {\bf G} = 0 \label{Max2} 
\end{eqnarray}
where the vector-matrix ${\bf s} = (s_x, s_y, s_z) = (s_1, s_2, s_3)$ is the vector with the three following components $(s_{i})_{kl} = -\imath e_{ikl}$, where $e_{ikl}$ is the 
absolute antisymmetric tensor. Note that the four Maxwell equations are now reduced to the two groups of two equations in each and one group contains only vector ${\bf F}$, while 
another group contains only vector ${\bf G}$. Moreover, each of the equations with the time derivative is similar to the corresponding Dirac equations for the spinor wave function. 
The total equation of the free electron field is a bi-spinor function, while the total wave function of the free-electromagnetic field is a bi-vector function. It is interesting to 
note that the 3$\times$3 matrixes $s_x, s_y, s_z$ in Eqs.(\ref{Max1}) - (\ref{Max2}) play the same role for photons as the Pauli matrixes $\frac12 \sigma_x, \frac12 \sigma_y, \frac12 
\sigma_z$ play for electrons. Therefore, they can be considered as spin matrixes. The commutation rules for the matrixes from these two groups of generators of transformations are 
similar: $s_i s_k - s_k s_i = \imath e_{ikl} s_l$ and $(\frac12 \sigma_i) (\frac12 \sigma_k) - (\frac12 \sigma_k) (\frac12 \sigma_i) = \imath e_{ikl} \frac12 \sigma_l$. For electrons 
the vector with the components ($\frac12 \sigma_x, \frac12 \sigma_y, \frac12 \sigma_z$) is the spin vector. Therefore, for the vector $s_x, s_y, s_z$ is the spin vector of a photon. 
The Casimir operator of the second order $C_2$ for this algebra equals 2, i.e. $C_2 = s (s + 1) = 2$ and we can say that the spin $s$ of a single photon equals unity. 

The main advantage of such a fomr of Maxwell equations is very simple and transparent formulas which describe behavior of the bi-vectors $({\bf F}, {\bf G})$ under Lorentz 
transformations, i.e. under rotations and velocity shifts. The explicit formulas take the form
\begin{eqnarray}
 {\bf F} \rightarrow \Bigl[ 1 + \frac{\imath}{4 \pi} {\bf s} \cdot \delta(\vec{\theta}) - \frac{1}{c} {\bf s} \cdot \delta {\bf v} \Bigr] {\bf F} \label{trans1} \\
 {\bf G} \rightarrow \Bigl[ 1 + \frac{\imath}{4 \pi} {\bf s} \cdot \delta(\vec{\theta}) - \frac{1}{c} {\bf s} \cdot \delta {\bf v} \Bigr] {\bf G} \label{trans2} 
\end{eqnarray}
These formulas for the Lorentz transformations of bi-vectors of the free electromagnetic field are very similar to the formulas for the Lorentz transformations
derived for the electron wave functions which is a bi-spinor $(\xi, \eta)$
\begin{eqnarray}
 \xi  \rightarrow \Bigl[ 1 + \frac{\imath}{8 \pi} \vec{\sigma} \cdot \delta(\vec{\theta}) - \frac{1}{2 c} \vec{\sigma} \cdot \delta {\bf v} \Bigr] \xi \label{trans11} \\
 \eta \rightarrow \Bigl[ 1 + \frac{\imath}{8 \pi} \vec{\sigma} \cdot \delta(\vec{\theta}) - \frac{1}{2 c} \vec{\sigma} \cdot \delta {\bf v} \Bigr] \eta \label{trans21} 
\end{eqnarray}
where $\frac12 \vec{\sigma} = \frac12 (\sigma_x, \sigma_y, \sigma_z)$ is the electron spin vector. In this case the Casimir operator $C_2$ equals $\frac{3}{4}$ and the electron spin 
equals $\frac12$. Here we do not want to discuss other properties of the Majorana representation \cite{Major} of Maxwell equations. Note only that this representation is very useful 
in application to some electromagnetic problems. In the next Section, we briefly discuss the so-called `symmetric form' of Maxwell equations.  

\section{On the symmetric form of Maxwell equations}

The ideas discussed in this Sections were originally stimulated by Dirac's research on magnetic monopole \cite{Dirac31}. It is a controversial matter which recently attracted a very 
substantial attention. Originally, my plan was to extend and publish this Section as a separate manuscript, but these days it is really hard to publish a manuscript, if its subject 
contradicts foundations of something (e.g., classical electrodynamics) known to everybody. On the other hand, our conclusions agree with a number of facts known from everyday life. 
Finally, I decided to write the text in the form which allows readers to make personal decisions about this subject.   

Let us consider the general Maxwell equations for the classical EM-field (see, e.g., \cite{Heitl})
\begin{eqnarray}
 && curl {\bf H} = \frac{1}{c} \frac{\partial {\bf E}}{\partial t} + \frac{4 \pi}{c} {\bf j}_e \; \; \; , \; \; \; 
 curl {\bf E} = - \frac{1}{c} \frac{\partial {\bf H}}{\partial t} \label{Maxwell} \\
 && div {\bf E} = 4 \pi \rho_e \; \; \; , \; \; \; div {\bf H} = 0 \nonumber
\end{eqnarray}
where $\rho_e$ and ${\bf j}_e$ are the electric charge density distribution and the current of electric charges. Note that in Eq.(\ref{Maxwell}) the $\rho_e$ is the true scalar 
and ${\bf j}_e$ is a true vector. The equations, Eqs.(\ref{Maxwell}) are written in the so-called non-symmetric form, since they contain, e.g., the electric current ${\bf j}_e$, 
but no analogous magnetic current ${\bf j}_m$. Their `manifestly symmetric' form of these equations is
\begin{eqnarray}
 && curl {\bf H} = \frac{1}{c} \frac{\partial {\bf E}}{\partial t} + \frac{4 \pi}{c} {\bf j}_e \; \; \; , \; \; \; 
 curl {\bf E} = - \frac{1}{c} \frac{\partial {\bf H}}{\partial t} + \frac{4 \pi}{c} {\bf j}_m \label{Maxwel1} \\
 && div {\bf E} = 4 \pi \rho_e \; \; \; , \; \; \; div {\bf H} = 4 \pi \rho_m \nonumber
\end{eqnarray}
where $\rho_m$ is the magnetic charge density distribution (pseudo-sclar) and ${\bf j}_m$ (pseudo-vector) is the current of magnetic charges. It is clear that the four quantities 
($\rho_m$ and three components of ${\bf j}_m$) form the four-vector (or four-pseudo-vector) which is properly transformed under the Lorentz transformation. In our `real' world we 
have no free magnetic charges. Not even a single stream (or current) of magnetic charges was ever observed. On the other hand, it can be another world where free magnetic charges 
and currents of such charges do exist, but the presence of free electric charges is impossible. It can be shown that events in these two worlds proceed absolutely independent, 
i.e. these events cannot affect each other, since the cross-sections between events which proceed in these two worlds are always equal zero. In the lowest order approximation upon 
$\alpha$ (the fine structure constant) interaction and/or communication between these two worlds can be produced by radiation, i.e. by regular electromagnetic waves. 

First, let us find Maxwell equations which govern all electromagnetic phenomena in that `alternative' (or magnetic) world. Assuming the absolute separation of the two worlds we 
can write from Eqs.(\ref{Maxwel1})
\begin{eqnarray}
 &&  curl {\bf H} = \frac{1}{c} \frac{\partial {\bf E}}{\partial t} \; \; \; , \; \; \; 
 curl {\bf E} = - \frac{1}{c} \frac{\partial {\bf H}}{\partial t} + \frac{4 \pi}{c} {\bf j}_m \label{Maxwel2} \\
 && div {\bf E} = 0 \; \; \; , \; \; \; div {\bf H} = 4 \pi \rho_m \nonumber
\end{eqnarray}
In other words, the electric field vector is now solenoidal (i.e. $curl {\bf E} = 0$), while the magnetic field has sources. Another interesting observation follows from 
Eqs.(\ref{Maxwel2}). If pseudo-scalar $\rho_m$ and pseudo-vector ${\bf j}_m$ equal zero identically, then Eqs.(\ref{Maxwel2}) coincide with Maxwell equations known for the free 
electromagnetic field. The same equations are correct in our `real' space. This means that we can register EM-waves which are coming from that `alternative' world. It works in 
the opposite way too: they can observe EM-waves which have been emitted in our space. Briefly, this means that two our worlds are complement to each other. Furthermore, these 
two worlds can be considered as the two separated parts of one United super-world. It is impotant that our (or real) world and ghost world can be located at the same place (a 
local piece of `our' three-dimensional space) and the `door' between these two worlds is the reflection in some `actual' mirror. Here by an `actual' mirror I mean a mirror which: 
(1) reflects all objects as a regular mirror, and (2) transforms all scalar, vectors and tensors into pseudo-scalars, pseudo-vectors and pseudo-tensors, respectively. An opposite 
transformation of `pseudo-'values into the `actual' values also takes place during such a reflection. The second point in this definition is crucial, since currently the word 
`reflection' in physics is overloaded with different meanings. 

As mentioned above the electric and magnetic worlds which are complement to each other. In general, these two worlds can be considered as the two independent components of one united 
super-space of events, or Super-World. This is of great interest for a large number of applications. For instance, for theology this means that the life and death are the two complementary 
forms of one super-life (or super-existence) and transformation between these two form is the reflection in the `actual' miror as defined above. Remarkably, that all facts known from the 
`old' religous points of view, e.g., `spiritual transitions from life to death and vise versa, i.e. from death to life', `spirit risen from the death', etc are supported by Maxwell 
equations in their symmetric form and can formally be described by these equations. The question about observation and registration of radiation which comes from the `magnetic' world (or 
`ghost' world) is very interesting. However, here one finds two questions which must be answered before such observations will be possible. First, right now we know almost nothing about 
frequencies and amplitudes of radiation which arives into our world from its `magnetic' counterpart. Very likeley, it has relatively low frequencies and small amplitudes. Second, we do 
not know the exact moment when any pulse of radiation will be emitted in the magnetic world. Therefore, it is hard to predict the moment of registration and the corresponding frequencies. 
Probably, a few `gifted' people can see such a radiation and respond to it, but any systematical, experimental study of radiation ariving from the magnetic world into our electric world is 
an extremely complex process. Despite a large number of unanswered questions at this moment we can only predict that old-fashioned Maxwell equations (without any modification) are the
appropriate, accurate and sufficient `tool' to provide communications between our real world and the `ghost' world.   

\section{On scalar electrodynamics}

As is well known the equations of electromagnetic field(s), or Maxwell equations, contain one polar vector ${\bf E}$ and one axial vector ${\bf H}$. All components 
of these two vectors are unknown functions of the spatial coordinates ${\bf r}$ and time $t$. From here one can conclude that an arbitrary electromagnetic field 
always has six independent (unknown) components which are the scalar components of these two vectors. However, this conclusion is not correct, since by using formulas 
known for the Lorentz transofrmations between two different inertial systems we can reduce the total number of independent components to four. It is clear that a 
possibility of such a reduction is closely related to the well known fact that there are two independent field invariants ${\bf E}^2 - {\bf H}^2$ (scalar, which up to 
the constant is the Lagrangian (or Lagrangian density) of the electromagnetic field) and ${\bf E} \cdot {\bf H}$ (pseudoscalar). This allows us to introduce the 
four-vector of field potentials $(\Phi, {\bf A})$ and re-write all Maxwell equations in terms of $\phi$ and ${\bf A}$. In the general form these equations are (in 
regular units):
\begin{eqnarray}
 \frac{4 \pi}{c} \rho {\bf v} &=& \frac{1}{c^2} \frac{\partial^2 {\bf A}}{\partial t^2} - \nabla^2 {\bf A} + grad \Bigl( div {\bf A} + 
 \frac{1}{c} \frac{\partial \Phi}{\partial t} \Bigr) \label{Meq1} \\ 
  4 \pi \rho &=& -\nabla^2 \Phi - \frac{1}{c} \frac{\partial}{\partial t} (div {\bf A}) \label{Meq2}
\end{eqnarray}
where the second equation can be re-written to the form
\begin{eqnarray}
  4 \pi \rho &=& \frac{1}{c^2} \frac{\partial^2 \Phi}{\partial t^2} - \nabla^2 \Phi - \frac{1}{c} \frac{\partial}{\partial t} \Bigl( div {\bf A} + 
  \frac{1}{c} \frac{\partial \Phi}{\partial t} \Bigr) \label{Meq3}
\end{eqnarray}
The vector ${\bf A}$ is defined by the differential equation $curl {\bf A} = {\bf H}$. It follows from here that the vector ${\bf A}$ is defined up to a gradient
of some scalar function, i.e. our equations must be invariant during the transformation: ${\bf A}^{\prime} \rightarrow {\bf A} + \nabla \Psi$. The choice of this 
function ($\Psi$) can be used to simplify the both equations, Eqs.(\ref{Meq1}) and (\ref{Meq3}). This is very well known gauge invariance (or gauge freedom) of the 
Maxwell equations. It is well described in numerous books on classical electrodynamics (see, e.g., \cite{Heitl} and \cite{LLE}). A freedom to chose different gauges 
is often used to solve actual problems in electrodynamics. Here we do not want to discuss it. Instead let us consider a slightly different approach which can be
very effective for many complex problems in electrodynamics. This approach is called the `scalar electrodynamics'.    

By analyzing equations Eqs.(\ref{Meq1}) - (\ref{Meq3}) one finds that to solve these equations we need to determine the four scalar functions, e.g., the scalar potential 
$\Phi$ and three components of the vector-potential ${\bf A} = (A_{x}, A_{y}, A_{z})$, or their linear combinations. This approach is absolutely equivalent to the use of 
one four-vector $(\Phi, {\bf A})$, but the use of non-covariant notations instead of one four-vector does not lead to any simplification in the general case. However, 
there is another approach which is based on the following theorem from Vector Calculus \cite{Kochin}. An arbitrary vector ${\bf a}$ is uniformly represented in the form 
\begin{eqnarray}
    {\bf a} = \phi \cdot grad \psi + grad \chi =  \phi  \nabla \psi + \nabla \chi \label{scal1}
\end{eqnarray}
where $\phi, \psi, \chi$ are the three scalar functions which depend upon three spatial coordinates ${\bf r}$ and time $t$. The proof of this theorem is relatively 
simple (see, e.g., \cite{Kochin}) and it leads to the following identity: $curl {\bf a} = grad \phi  \times  grad \psi = \nabla \phi \times \nabla \psi$. The 
expression for the $div {\bf a} = \nabla {\bf a}$ is slightly more complex: $div {\bf a} = grad \phi  \cdot  grad \psi +  \phi \Delta \psi + \Delta \chi = \nabla \phi  
\cdot  \nabla \psi +  \phi \Delta \psi + \Delta \chi$. If we can chose the functions $\psi$ and $\chi$ as the solutions of the Laplace equations, i.e. $\Delta \psi = 
0$ and $\Delta \chi = 0$ (i.e. these two functions are the harmonic functions), then from the last equation one finds  $div {\bf a} = grad \phi  \cdot  grad \psi = 
\nabla \phi \cdot \nabla \psi$. In classical electrodynamics we can always represent the vector-potential ${\bf A}$ in the form of Eq.(\ref{scal1}). Then the solution 
of the incident problem is reduced to the derivation of the corresponding equations for the three scalars $\phi, \psi, \chi$ in Eq.(\ref{scal1}) and scalar-potential 
$\Phi$ form the four-vector $(\Phi, {\bf A})$. An obvious advantage of this method follows from the fact that we can chose three functions $\phi, \psi, \chi$ 
step-by-step and by using the known boundary and initial conditions. For many problems it provides crucial simplifications of arising equations and allows one to find 
the explicit solutions. However, in this approach Maxwell equations become a system of the non-linear equations. For theoretical development of the classical/quantum 
electrodynamics this approach (based on Eq.(\ref{scal1})) has never been used.

\section{Conclusion}

Thus, we have applied the methods of constraint dynamics developed by Dirac \cite{Dirac0}, \cite{Dirac1} to derive the Hamiltonian of the free electromagnetic field. This 
Hamiltonain and arising primary and secondary constraints are used to derive the corresponding Schr\"{o}dinger equation for the free electromagnetic field. One of the 
advantages of this method is the absence of any scalar and/or longitudinal photons. Both scalar and longitudinal photons arise in standard QED, since without them this 
theory cannot be considered as being a closed, relativistic procedure. However, the presence of such photons in the expression for the electric and magnetic fields makes all 
QED calculations extremely difficult. Furthermore, initially in QED the physical (or internal) reasons for the appearance of scalar and longitudinal photons was not 
clear. Fermi proposed to exclude all such `non-physical' photons by using re-definition of the field wave functions \cite{Fer}. At the same time a number of other ideas and 
recipes were proposed which lead to complete exclusion of the scalar and/or longitudinal photons from QED calculations. Only after development of the constrained dynamics 
by Dirac it became clear that the original Fermi's idea is essentially correct. Based on Dirac's methods we have develop a new approach to perform QED calculations which are 
correct at each step. This approach will be described elsewhere.

We also briefly discuss questions related with the dynamical symmetry of the free electromagnetic field, Majorana form of the Maxwell equations, communications between our 
(electric) and ghost (or magnetic) worlds and sclalar approach to electrodynamics which is based on the use of four scalar functions only. It should be mentioned that Maxwell 
theory of electromagnetic equations is one of the oldest (150 years old!) physical theories which is still in active use in many areas of modern science and technology. 
Nevertheless we are still far away from that moment when we can say that we know everything about Maxwell equations and predictions which follow from these equations at 
different experimental conditions. This indicates clearly that Maxwell theory of electromagnetic filed(s) is healthy and it is still a subject of intense theoretical and 
experimnetal development.  

\begin{center}
 {\Large Appendix.} 
\end{center}

Here we discuss the role of the secondary constraint $div {\bf E} = 0$ in Dirac's electrodynamics. Recently, in many books and textbooks it became a tradition 
to consider all primary and secondary constraints for the free electromagnetic field as some secondary conditions which play a non-significant role (in contrast with 
the Hamiltonian equations) for the field itself. From my point such a view is absolutely wrong and may lead to serious mistakes, if it applies to other fields. Even 
for the free electromagnetic field the constraint $div {\bf E} = 0$ allows one to predict many important details of its propagations. Let us discuss this problem here. 
First, note that the constraint $div {\bf E} \mid \Psi \rangle = 0$ exactly coincides with one of the field equations (or Maxwell equations). There is no easy way to 
derive this equation by using the Hamiltonian of the free electromagnetic field. This condition means that no new (non-zero) electric charge can be created during any 
possible time-evolution of the free electromagnetic field in our three-dimensional space. In addition to this, the condition $div {\bf E} \mid \Psi \rangle = 0$ 
substantially determines the actual shape and time-evolution of the free electromagnetic field. Indeed, let us consider the formula for the divergence of the vector 
${\bf E}$ in spherical coordinates $(r, \theta, \phi)$ is
\begin{equation}
  div {\bf E} = \frac{1}{r^2} \frac{\partial (r^2 E_r)}{\partial r} + \frac{1}{r \sin\theta} \frac{\partial (\sin\theta E_{\theta})}{\partial 
 \theta} + \frac{1}{r \sin\theta} \frac{\partial (\sin\theta E_{\phi})}{\partial \phi} = 0 \; \; \; \label{eq50}
\end{equation}
where $E_r, E_{\theta}$ and $E_{\phi}$ are the spherical components of the ${\bf E}$ vector (the vector of the electric field intensity). Let us discuss possible choices 
of the spherical components of the vector ${\bf E} = ( E_r, E_{\theta}, E_{\phi})$ which will automatically lead to the identity $div {\bf E} = 0$. To obey the condition, 
Eq.(\ref{eq50}), the radial component $E_r$ of the vector ${\bf E}$ must be a very special function of $r$. The dependence of the $E_r$ component upon $r$ is general and 
it is crucially important for the whole electrodynamics in three-dimensional space. From the condition $\frac{\partial (r^2 E_r)}{\partial r} = 0$ one finds that at large 
$r$ the electric field intensity ${\bf E}$ decreases as $r^{-2}$, i.e. $E_r \simeq \frac{C}{r^2}$, where $C$ is some numerical constant. It can be shown that the same 
conclusion is true for the magnetic field intensity ${\bf H}$. Such a radial dependence at large $r$ is the known general property of the free electromagnetic field which 
propagates in three-dimensional space. Analogous conclusion about angular dependence of the $E_{\theta}$ and $E_{\phi}$ components (e.g. $E_{\theta} = \frac{F(r, 
\phi)}{\sin \theta}$ and $E_{\phi} = G(r, \theta)$, where $F$ and $G$ are the arbitrary (regular) functions of two arguments) cannot be considered as universal and general. 

\begin{center}
       {\bf Acknowledgments}
\end{center}

I am grateful to D.G.C. (Gerry) McKeon from the University of Western Ontario, London, Ontario, CANADA) for helpful discussions and inspiration.

\end{document}